\documentclass[11pt,twocolumn,italian,english]{article}
\usepackage[T1]{fontenc}
\usepackage[latin9]{inputenc}
\usepackage[landscape,a4paper]{geometry}
\usepackage{fancyhdr}
\usepackage{color}
\usepackage{babel}
\usepackage{amstext}
\usepackage{graphicx}
\usepackage[unicode=true,pdfusetitle,
 bookmarks=true,bookmarksnumbered=false,bookmarksopen=false,
 breaklinks=true,pdfborder={0 0 1},backref=false,colorlinks=true]
 {hyperref}
\hypersetup{
 linktocpage=true,citecolor=viola, linkcolor=arancio,urlcolor=viola}

\makeatletter

\newcommand{\noun}[1]{\textsc{#1}}

\@ifundefined{date}{}{\date{}}
\usepackage{multicol}

\makeatletter
\let\ps@plain\ps@fancy   
\makeatother
\definecolor{rossos}{cmyk}{0,1,1,0.55}
\definecolor{bluscuro}{rgb}{0.15, 0.2, .85}
\definecolor{grigioaltocontrasto}{rgb}{0.37, 0.37, 0.37}
\definecolor{bluchiaro}{cmyk}{1,.3,0.,0.1}
\definecolor{viola}{cmyk}{0.65,0.70,0,0}
\definecolor{arancio}{cmyk}{0.10,0.60,1,0}
\usepackage[usenames,dvipsnames]{xcolor}
\usepackage[font=small,labelfont={it,bf}]{caption}

\makeatother

\begin{document}

\title{\noun{Energy peaks: a high energy physics outlook}}

\author{Roberto Franceschini\thanks{roberto.franceschini@uniroma3.it}\\
\foreignlanguage{italian}{{\small{}Dipartimento di Matematica e Fisica,
Università degli Studi Roma Tre and INFN, sezione di Roma Tre, I-00146
Rome, Italy}}}
\maketitle
\begin{abstract}
{\footnotesize{}Energy distributions of decay products carry information
on the kinematics of the decay in ways that are at the same time straightforward
and quite hidden. I will review these properties and discuss their
early historical applications as well as more recent ones in the context
of }\emph{\footnotesize{}i}{\footnotesize{}) methods for the measurement
of masses of new physics particle with semi-invisible decays, }\emph{\footnotesize{}ii}{\footnotesize{})
the characterization of Dark Matter particles produced at colliders,
}\emph{\footnotesize{}iii}{\footnotesize{}) precision mass measurements
of Standard Model particles, in particular of the top quark. Finally
I will give an outlook of further developments and applications of
energy peaks method for high energy physics at colliders and beyond.}{\footnotesize \par}
\end{abstract}
{\footnotesize{}\tableofcontents{}}{\footnotesize \par}

\section{Introduction}

Energy is probably the most frequently used quantity to characterize
a system in physics. Despite this, for very good reasons, in certain
types of (high energy physics) experiments energy is not often used
to present the data. In these pages I will review why this is the
case and what can be gained in these experiments  using energy as
a primary quantity to look at the data.

For a system of mass $m$ and momentum $p$ the total energy is $E=\sqrt{p^{2}+m^{2}}$.
In high energy physics is often the case that the mass of the system,
typically a particle, can be neglected with respect to the momentum
of this particles, that is to say in high energy physics we often
deal with objects close to their ultra-relativistic regime. In these
circumstances energy and momentum become synonymous as $E=p$. This
is a first reason why theoretical as well experimental discussions
in high energy physics are often carried out in terms of momentum.
Further to it, in collider experiments such as those at the Large
Hadron Collider at CERN the energy is so high that the colliding protons
do not interact as single particles - they reveal their constituent
particles each carrying an unknown fraction of the proton energy.
As a consequence, collisions between proton constituents happen at
unknown energies, making difficult to use energy to characterize these
collisions. As a matter of fact in high energy physics experiments
it is often exploited the fact that the colliding particles and their
constituents travel at high speed along a well specified direction,
so that it is safe to assume that any motion of the collision products
in the perpendicular directions is the result of the interactions
happened in the collision. This line of thinking lead in the past
decades to development of a large body of research \cite{Barr:2011ao}
about how to study the details of particles collisions looking only
at the momentum in the direction perpendicular to the direction of
the colliding protons. The observables that use only information from
the momentum in the perpendicular direction are called \emph{transverse},
as opposed to \emph{longitudinal}. For momentum this division results
in the definition $\vec{p}=\vec{p}_{T}+\vec{p}_{L}$, where traditionally
$p_{L}=p_{z}$ and $p_{T}=\sqrt{p_{x}^{2}+p_{y}^{2}}$ in cartesian
coordinates. In these coordinates energy is given by $E=\sqrt{p_{x}^{2}+p_{y}^{2}+p_{z}^{2}}$
, which is invariant under spatial rotations of the reference axes,
but transforms under Lorentz transformations as 
\[
E^{\prime}=\gamma E+\gamma\sqrt{1-\gamma^{2}}\cos\theta|\vec{p}|\,,
\]
for a transformation of velocity parameter $\vec{\beta}$ directed
with an angle $\theta$ with respect to the direction of the momentum
$\vec{p}$ and resulting in a Lorentz factor $\gamma=1/\sqrt{1-|\vec{\beta}|^{2}}$.
Under the same transformation momentum components $p_{i}$ transform
as 
\[
p_{i}^{\prime}=\gamma p_{i}-\gamma\beta_{i}E\,,
\]
therefore if the velocity of the Lorentz transformation is orthogonal
to $\vec{p}$ the transformation has no effect on $\vec{p}$. This
implies that a boost directed along the $z$ axis (that is the beam
axis) will leave unchanged the \emph{transverse} momentum $p_{T}$.
Clearly the invariance of this quantity is another good reason for
using \emph{transverse} momentum for many discussion of physics at
particle colliders.

Nevertheless, restricting experiments to just use \emph{transverse}
observables effectively throws away part of the information on the
collisions that the experiments set out to study. Furthermore there
are cases in which combining \emph{transverse }and \emph{longitudinal}
observables allows to recover important kinematical properties that
we are going to review in the following Sections for the case the
energy.

These properties are somewhat standard knowledge of the cosmic ray
physics practitioners and we will see how they have been rediscovered
and extended in recent years in high energy particle physics and what
prospect they still have for future applications.

\section{History and cosmic rays peaks \label{sec:History}}

The observation of particles hitting Earth from outer space has traditionally
been a carrier of discoveries and surprise in particle physics. Indeed
many particles have been discovered in the cosmic radiation before
they could be produced in laboratories with accelerators. One question
that was in debate in the early 1950s in the particle physics community
was the composition of the natural radiation that can be measured
in the atmosphere with growing intensity at higher altitudes, what
we call today cosmic rays. This radiation is mostly the result of
interactions of primary protons hitting the upper atmosphere layer
giving rise to a ``shower'' of cosmic ray secondary products that
proceed further towards ground level and beyond. In the early 1950s
one experiment was carried out to identify the origin of photons that
appeared in the cosmic radiation and concluded that these photons
were the trace of the presence of $\pi^{0}$ in the cosmic radiation.
The $\pi^{0}$, in modern view, are expected to be abundantly produced
when the atmosphere is hit by protons, but their presence was not
experimentally proven until the measurement of Carlson \cite{Carlson}.
This experiment identified $\pi^{0}\to\gamma\gamma$ decay by measuring
the spectrum of $\gamma$ and observing that it had a peak at $m_{\pi^{0}}/2$,
which was expected from relativistic kinematics for the distribution
of the energy of massless decay products from a heavy particle decay.

The argument is very simple and short and can be found in Ref.~\cite{Agashe:2013sw}
as well as in standard textbooks \cite{Gaisser:219622} of cosmic
rays physics. It essentially states that if we observe all the decay
products of a scalar particle that decays in to \emph{massless} particles,
a process we denote $M\to ab$, the distribution of the energy of
either of the decay products, say $a$, has a peak at $m_{M}/2$ irrespective
of the distribution of boosts of the decaying particle $M$ in the
frame in which we carry out the measurement. Given its simplicity
it is not surprising that this observation has re-surfaced the literature
on cosmic rays in more than one occasion, sometimes with extensions
of the original argument and new applications to new domains \cite{kopylov}\cite{1971NASSP.249.....S}.

In modern problems of cosmic rays physics, however, it seems that
this property of energy distributions is of limited use. For the study
of complicated air-showers resulting from protons hitting the atmosphere
this type of characterization of the decay product of a particle are
insufficient to achieve interesting results and more complex modeling
of these phenomena is necessary. Still, the of $\pi^{0}$ characteristic
peak in the photon spectrum has allowed the FERMI collaboration to
claim the identification of $\pi^{0}$ in the gamma rays from sources
known to be remnants of supernova explosions \cite{Ackermann:2013oq}
- quite an achievement to find out particles identity observing their
decay products from few thousand light-year distance! This example
well represents the possible domain of application of energy peaks,
as it shows how, despite the large distance, despite the uncertain
conditions of production of the $\pi^{0}$ in the supernova remnant
environment, and despite the propagation of the $\gamma$ from the
source to us, the energy spectrum of these photons still carries enough
information to allow to identify its parent particle. Further to it,
we are identifying the parent particle observing just one of the photons
decay product at a time. This is remarkable because if we had measured
energy and direction of both photons for each $\pi^{0}\to\gamma\gamma$
decay we would know from four-momentum conservation that the parent
particle was a pion, we would simply observe that 
\begin{equation}
\left(p_{\gamma_{1}}+p_{\gamma_{2}}\right)^{2}=m_{\pi^{0}}^{2}\,.\label{eq:pi2yy}
\end{equation}
 On the contrary, looking the energy peak distribution enables us
to make a statement on the origin of the photons despite our complete
ignorance on the properties of the other photon produced in the decay. 

\section{Recent Applications in high energy physics}

The property of the peaks of energy distributions well known for spinless
particles such as the neutral pion has been recently extended to the
case of particles with spin \cite{Agashe:2013sw}. The key observation
that allows to extend the result to particles with spin is that when
decays are observed from a collection of particles that populates
evenly all the polarization states, this collection of particles is
statistically equivalent to a scalar. Therefore if one observes decays
of a massive fermions belonging to a collection made equally of left-handed
and right-handed massive fermions the energy distribution of the decay
products is guaranteed to have the same peak as if the decaying fermions
were scalars. Therefore for a fermion parent particle $F$ decaying
into two massless particles
\[
F\to ab
\]
the energy spectrum of either decay products has a peak at $m_{F}/2$
irrespective of the distribution of boosts of the decaying particle
$F$ in the frame in which we carry out the measurement. 

This observation enables the use of energy peak distributions in a
completely new set of problems, especially in high energy particle
physics experiments, where fundamental particles with spin, \emph{e.g.}
the quarks and leptons of the Standard Model, are studied.

In the following we will deal with decays in which both the final
state particles have a non-vanishing mass, in such case the energy
peak for the particle $b$ is predicted to arise at

\begin{equation}
E_{b}^{*}=\frac{m_{F}^{2}-m_{a}^{2}+m_{b}^{2}}{2m_{F}}\,.\label{eq:master-two-body}
\end{equation}
This is the master formula that we will use for all the applications
to two-body decays. 

\section{New physics mass measurements}

As highlighted in the previous section \ref{sec:History} about the
historical uses of peaks of energy distributions, the presence of
a peak in the energy distribution of a decay product is a consequence
of general kinematical arguments. We have also highlighted how striking
it is to be able to say anything about cosmic gamma rays that (\emph{i})
have been produced in an environment of which we know so little details,
and (\emph{ii}) have traveled very long distances in a medium of rather
unknown properties. The condition of uncertainty in which energy peaks
properties have found successful application described above make
them a very interesting tool for the study of \emph{new} physics at
particle colliders. In fact when approaching a newly discovered particle
very little is know about it and is very useful to have tools to study
its properties making very minimal assumptions. For this reason the
first set of applications that have been studied in high energy physics
for the application of properties of energy peaks have to do with
the characterization of new physics, in particular putative new particles
to be discovered in collider experiments.

A classic problem one faces when a new particle is discovered is to
measure the mass of the new state. In most cases of particles discovered
in the past decades this problem was almost the same as that of discovering
the new particle itself. For example, observing a new peak in the
ratio $R=\sigma(e^{+}e^{-}\to hadrons)/\sigma(e^{+}e^{-}\to\mu^{+}\mu^{-})$
when the $e^{+}$ and $e^{-}$ beams reaches a new record high energy
automatically means a new particle has been produced and its mass
be around the center of mass energy of the beams, $\sqrt{s}=2E_{e^{+}}$
if the collision is a classic symmetric beam-beam collision. However,
in many new physics scenarios studied in nowadays experiments at particle
colliders such simple strategies are not applicable. A first example
of particle discovery in which the mass of the new particle is not
so obviously known are the discovery of the $W$ boson and that of
the top quark. These particles have been discovered observing decays
into final states containing invisible particles, \emph{e.g.} $W^{\pm}\to\ell^{\pm}\nu$
the decay of a $W$ boson into a charged lepton and a neutrino. The
neutrino, being a weakly interacting particle, is \emph{not detectable}
in the high energy experiments. Therefore we cannot use conservation
of four-momentum as in eq.(\ref{eq:pi2yy}) because one of the two
particles that emerges from the decay of the $W$ boson is not experimentally
accessible. The measurement of masses of particles that decay into
final states containing invisible particles (\emph{e.g.} neutrinos)
is one of the classic problems in mass measurement practice that can
receive new inputs from the use of energy peaks properties and we
will discuss examples of this kind at length.

When invisible particles are involved in the decay, as in the decay
of the $W$ boson, we are forced to find ``ad-hoc'' solutions to
measure the mass of the newly discovered particle. For the $W$ boson
the commonly adopted mass measurement method exploits the fact that
$W$ bosons are produced at hadron colliders in a very simple reaction
\begin{equation}
pp\to W+X\,,\label{eq:DYW}
\end{equation}
where $X$ denotes remnants of the protons that collided. In this
situation it is possible to apply the properties of transverse momentum
distribution know as the jacobian peak and further improve on these
results using a special kinematical variables called transverse mass~\cite{PhysRevLett.50.1738}.
Very good results have been obtained pursuing methods based on jacobian
peaks and transverse mass measurements. Nevertheless it is important
to remark that these methods rely on the knowledge of the details
of the production reaction eq.(\ref{eq:DYW}) and are indeed starting
to show limitations on the ultimate precision attainable in these
measurements. These limitations can be ultimately be traced back to
the fact that dynamics of the Standard Model of particles physics,
on top of simple kinematics, enters in the formulation of the methods
to measure the $W$ boson mass. For instance we have to know whether
or not there is a subdominant production mechanism which produces
pairs of $W$ bosons, or which quarks inside the proton collide to
produce a $W$ boson, as well as their momentum distribution. This
example should convince the reader that good strategies can be devised
on case-by-case basis when we know well enough the properties of the
particle subject of mass measurement, provided that we know all the
necessary inputs with sufficient accuracy. However without an accurate
knowledge of the necessary inputs, or in absence of a prevailing and
motivated underlying characterization of the particle, \emph{e.g.}
as a state foreseen in a given model, it is hard to make progress
in measuring particle masses when invisible final state particles
are involved. 

Given the vast landscape of possibilities for models of new physics
that have been conceived in the past decades it is of greatest importance
to be able to approach the problem of mass measurement with a minimal
amount of assumptions. The job of the high energy phenomenologist
is therefore to find strategies with an optimal balance between the
achievable precision on the mass measurement and the amount of assumptions
to be made to carry out the measurement. Clearly, as more data flows
in on the newly discovered particle, more and more assumption may
become justified and the point of balance in the determination of
a particle mass can switch more towards methods that use more dynamics,
but it remains a very important task to measure masses when these
assumptions are not yet fully justified - in this task is where we
expect energy peaks techniques to be very useful. 

In order to exemplify possible uses of energy distribution and peaks
identification in these distribution we recall a few applications
that have been studied in detail.

In Ref. \cite{Agashe:2013ff} the properties of energy peaks have
been applied to two body decays of supersymmetric particles, putative
new particles that arise in well motivated scenarios of new physics.
The case studied was the production of gluinos, denotes as $\tilde{g}$
as they are partners of the gluon particle $g$, that are produced
by strong interactions in 
\begin{equation}
pp\to\tilde{g}\tilde{g}\label{eq:guinopair}
\end{equation}
 and further decay first in a two-body decay 
\begin{equation}
\tilde{g}\to b\tilde{b}\label{eq:firststepgluino}
\end{equation}
and then the sbottom particle $\tilde{b}$, the partner of the bottom
quark particle $b$, decays into two bodies
\begin{equation}
\tilde{b}\to b\chi^{0}\label{eq:secondstepgluino}
\end{equation}
yielding overall a cascade decay 
\begin{equation}
\tilde{g}\to b\tilde{b}\to bb\chi^{0}\,,\label{eq:cascade}
\end{equation}
that is the decay of a single gluino into two bottom quarks and a
neutral particle, denoted by $\chi^{0}$, that is know as neutralino.
The neutralino is, like a neutrino, charged only under weak interactions,
therefore it does not leave any direct trace in detectors. For this
reason we face, as for the case of the $W$ boson illustrated above,
a decay into a semi-invisible final state, which poses serious challenges
for a mass measurement.

Considering that two gluinos are produced in each reaction eq.(\ref{eq:guinopair}),
overall we have
\[
pp\to\tilde{g}\tilde{g}\to bbbb\chi^{0}\chi^{0}\,,
\]
that is a quite busy reaction with two invisible particles, the two
$\chi^{0}$, and four identical particles, the four $b$ quarks, each
of the $b$ quarks giving rise to essentially indistinguishable signals
in the part of the detector that they will hit. Clearly, measuring
the gluino mass in such a reaction is quite challenging in many respects.
In fact, even if the $\chi^{0}$ particles were directly observables
in the detector, such a busy final state would pose a challenge in
understanding which particles come from one gluino and which form
the other gluino. In other words if one wants to use simple four-momentum
conservation as in the case of pion decay in eq.(\ref{eq:pi2yy})
one has first to single out the particles arising from each individual
gluino. On top of this, there is the extra difficulty $\chi^{0}$
are not measurable at all in our detectors. Therefore if one wants
to try to measure the mass with any strategy similar spirit to the
use of four-momentum conservation there are two big challenges to
be faced: \emph{i}) four-momentum is carried away by invisible particles
\emph{ii}) visible particles can be divided in subset belonging to
each gluino in a factorial number of ways~\footnote{At variance with the case of $W$ bosons considered above, here we
have the production of a pair of gluinos, so the ``ad-hoc'' solution
based on transverse mass and jacobian peaks adopted for singly produced
particles as the $W$ boson would not work here. Extensions of these
ideas have been attempted and resulted in the formulation of generalized
transverse mass and in particular of the s-transverse mass $m_{T2}$~\cite{Barr:2003fj,Lester:1999et},
but we will not discuss these ideas here. }. Clearly a new strategy for the gluino mass measurement might alleviate
significantly these obstructions - energy peaks measurements take
care of both these issues. 

The key idea of using energy peaks properties is to realize that information
about the mass difference between particles involved in each two-body
decay is carried by the peak position of the energy distribution.
In the example of gluino decay above we have 
\begin{equation}
E_{b}^{*}=\frac{m_{\tilde{g}}^{2}-m_{\tilde{b}}^{2}+m_{b}^{2}}{2m_{\tilde{g}}}\simeq\frac{m_{\tilde{g}}^{2}-m_{\tilde{b}}^{2}}{2m_{\tilde{g}}}\label{eq:Estepone}
\end{equation}
for the decay in eq.(\ref{eq:firststepgluino}) and 
\begin{equation}
E_{b}^{*}=\frac{m_{\tilde{b}}^{2}-m_{\chi}^{2}+m_{b}^{2}}{2m_{\tilde{b}}}\simeq\frac{m_{\tilde{b}}^{2}-m_{\chi}^{2}}{2m_{\tilde{b}}}\label{eq:Esteptwo}
\end{equation}
for the second step of the decay chain given in eq.(\ref{eq:secondstepgluino}),
where in both cases one may neglect the $b$ quark mass. Measuring
the two values of $E_{b}^{*}$ in energy spectrum of $b$ quarks can
therefore yield information on two masses involved in the problem,
\emph{e.g.} $m_{\tilde{g}}$ and $m_{\tilde{b}}$ if one assumes a
value for $m_{\chi}$ or evidence suggests it is negligible. Strikingly
the information about the two masses comes from the measurement of
a quantity that is directly accessible without any assumption or knowledge
about the invisible particle $\chi$. Furthermore these relations
hold for any process in which gluino is produced by strong interactions,
so if we had a mixture of $pp\to\tilde{g}\tilde{b}$ and $pp\to\tilde{g}\tilde{g}$,
which due to the complex collision environment at the Large Hadron
Collider might be look-alike processes, we can use these relations
to gather information on these two masses even in presence of such
a nuisance. 

\begin{figure}
\includegraphics[width=0.99\linewidth]{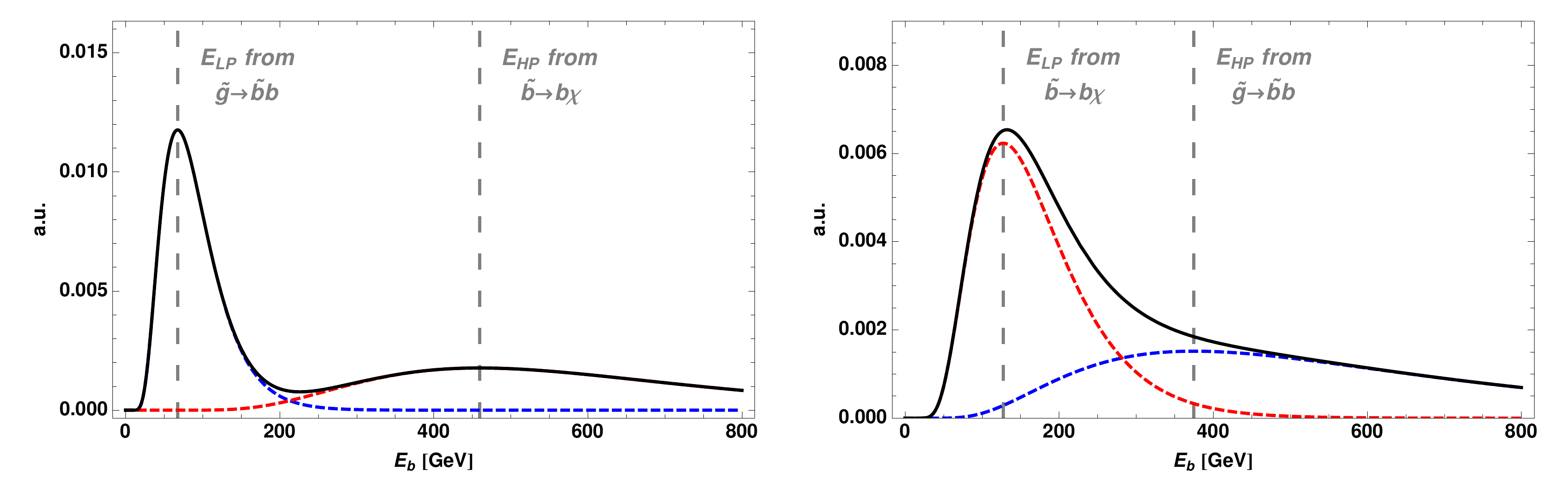}\caption{\label{fig:Two-examples-of-spectrum}Two examples of energy spectrum
that we can expect to measure from a two-step decay chain. The ``low
peak'' at $E_{LP}$ and the ``high peak'' at $E_{HP}$ are arising
from different steps of the decay chain in the two cases. Furthermore
the left panel spectrum has two well separated peaks, while the right
panel one has two peaks that tend to overlap.}

\end{figure}

Last but not least, attempting to measure masses with these single
particles relations has the advantage to not require us to identify
which $b$ quark is coming from each step of the decay. In general
each step of the decay process will give rise to a characteristic
spectrum and we will observe the sum of all the spectra coming from
all the decay steps. For our two-step decay chain we expect to have
spectra similar to one of the two distinctive cases shown in Figure~\ref{fig:Two-examples-of-spectrum}.
The right panel shows the spectrum of $b$ quark energies that one
expects when the two mass differences involved in are similar, while
the left panel shows the spectrum that we expect when the two mass
differences are significantly different. In the latter case it is
possible to attempt to extract from the spectrum the two values of
the energy peaks relation on the left-hand side of eq.(\ref{eq:Estepone})
and (\ref{eq:Esteptwo}) analyzing separately the ranges of energy,
one for each peak. An example of this type is described at length
in Ref. \cite{Agashe:2013ff}. 

\begin{figure}

\includegraphics[width=0.49\linewidth]{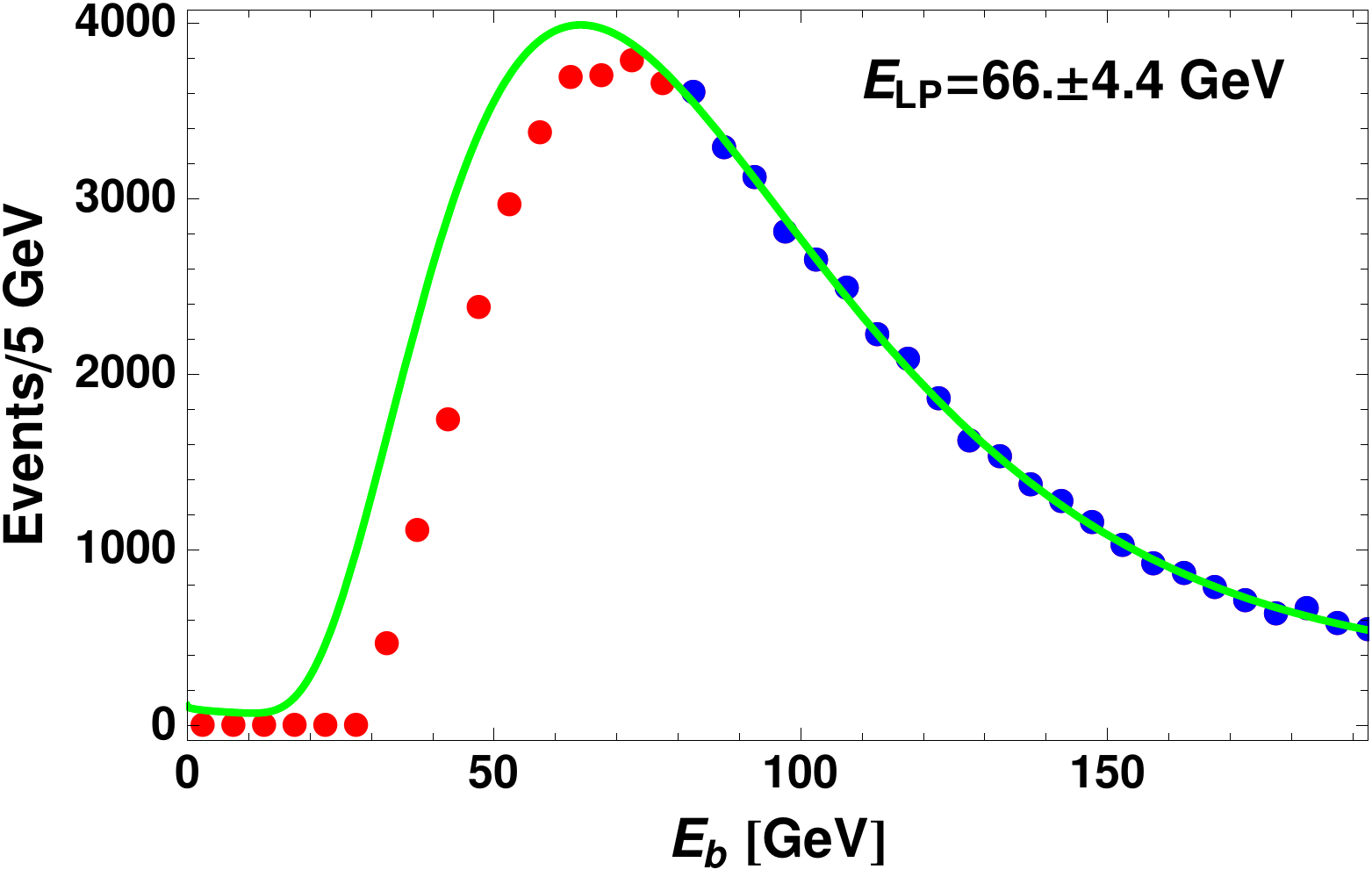}\includegraphics[width=0.49\linewidth]{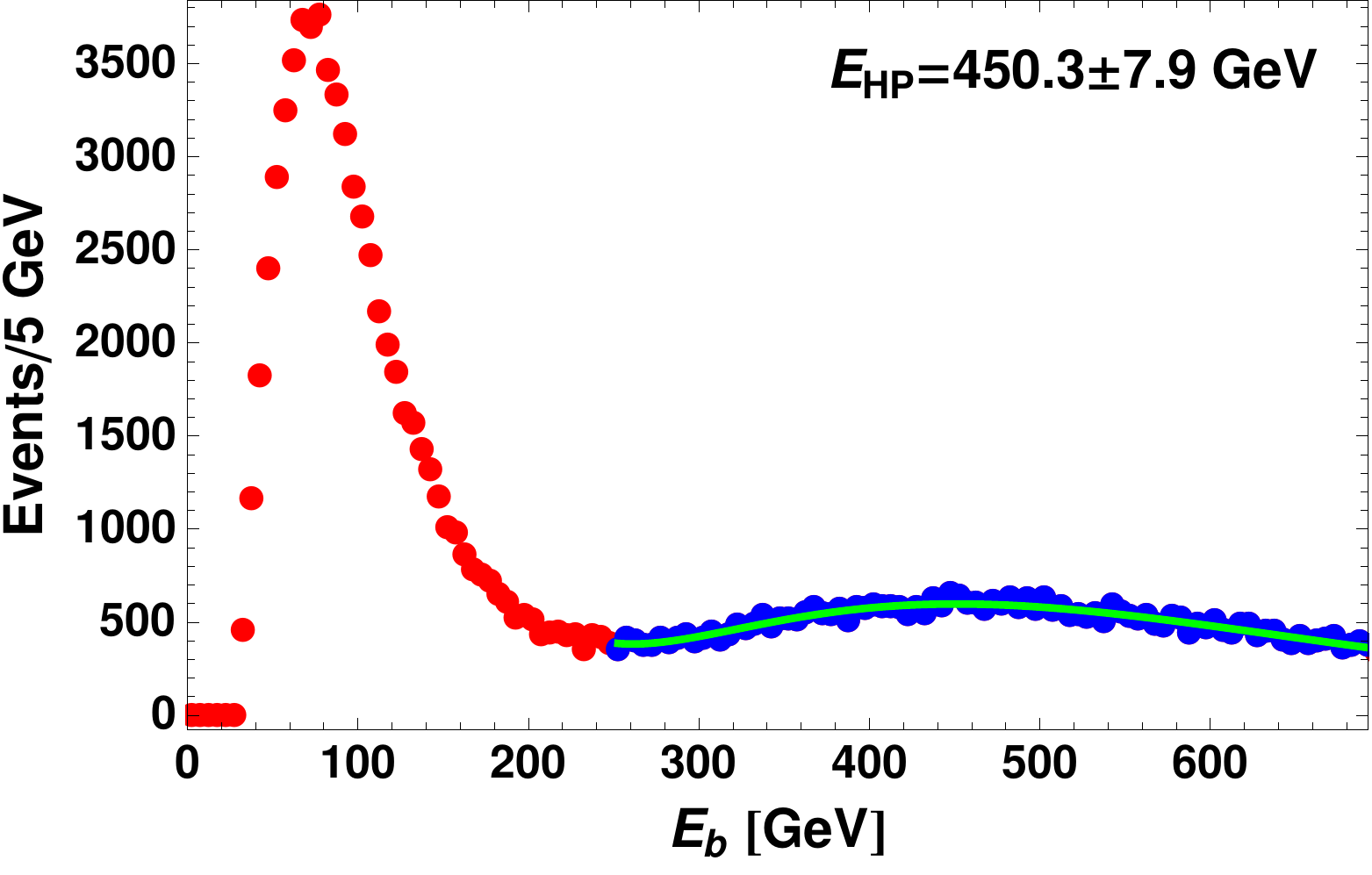}

\caption{\label{fig:Fits-separated}Fits of energy peaks an energy spectrum
containing two well separated peaks.}

\end{figure}

Here we report the result of the extraction of the two peaks as shown
in Figure~\ref{fig:Fits-separated}. The energy spectrum peak values
are identified through a fit in which the data is modeled by suitably
chosen function. This function has two parameters, one for the peak
position and one for the width of the peak shape. As explained in
Ref.~\cite{Agashe:2013sw}, from prime principle argument one can
see that this function has to depend on the combination $E^{*}/E+E/E^{*}$
, rather than being a simple function of $E$. In this example, and
many others, it has been found that data around the peak region can
be fit by 
\begin{equation}
A\cdot\exp\left\{ w\left(\frac{E^{*}}{E}+\frac{E}{E^{*}}\right)\right\} \,,\label{eq:fitfunction}
\end{equation}
where $E^{*}$ is the peak position, \emph{i.e.} the parameter of
interest for the mass measurement, $w$ fixes the width of the peak
shape, and $A$ is a normalization factor. The translation of the
peak fit results in actual mass measurement depends on the degree
of knowledge one assumes on $m_{\chi}$ as well as if one wants to
use further information on the masses that can be gathered in events
in which new physics particles undergo a cascade decay. As described
in detail in Ref. \cite{Agashe:2013ff} for the example at hand one
could exploit some features of the end-point of the distribution of
the invariant mass $m_{bb}$~\cite{Hinchliffe:1999ve}. All in all
Ref. \cite{Agashe:2013ff} finds that an extraction of $m_{\tilde{g}}$
and $m_{\tilde{b}}$ with precision better than 10\% is possible with
the energy peak method and can be improved to few percent precision
if additional information is used.

\begin{figure}
\begin{centering}
\includegraphics[width=0.49\linewidth]{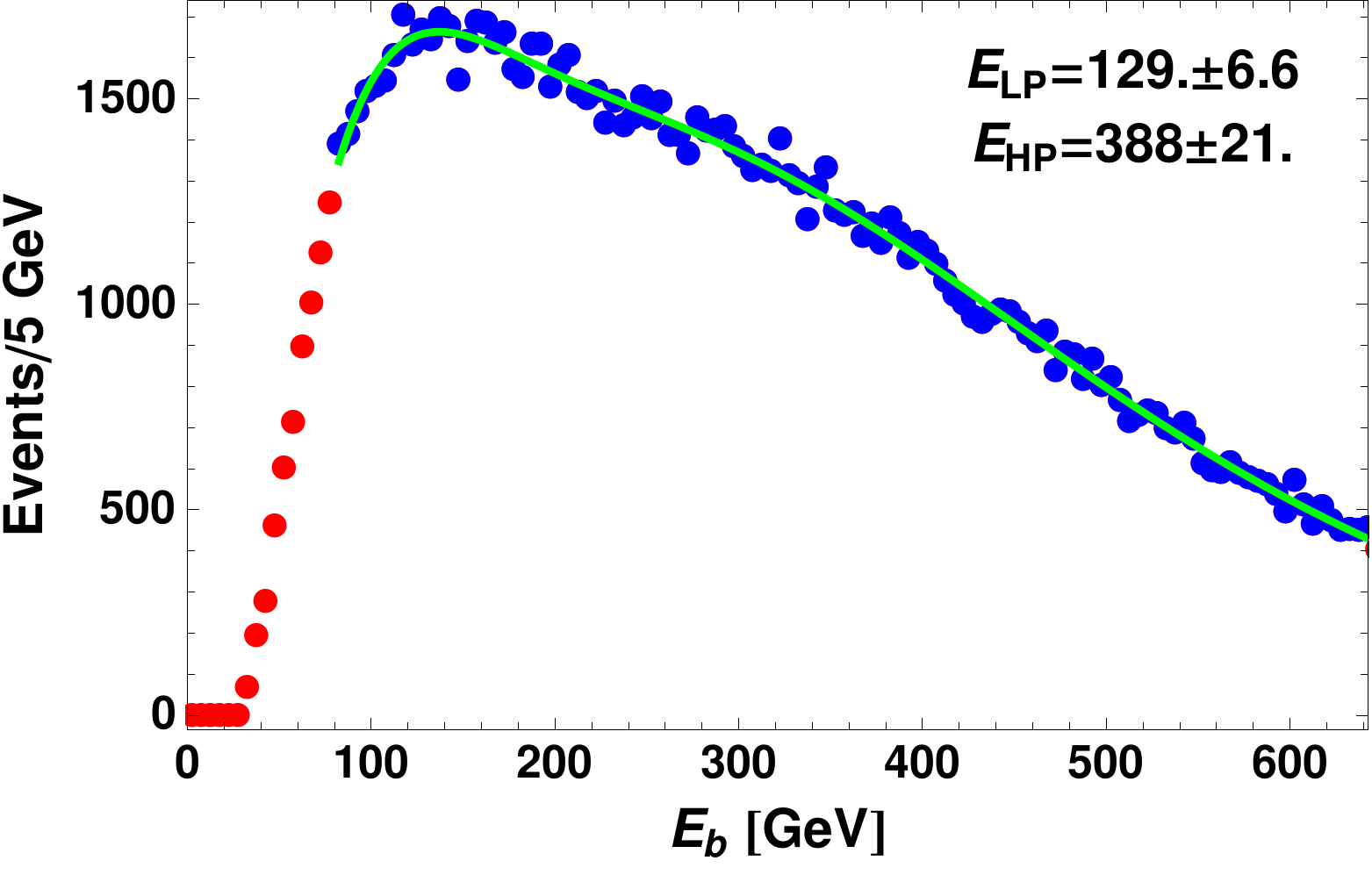}
\par\end{centering}
\caption{\label{fig:Fits-overlapping}Fits of energy peaks an energy spectrum
containing two overlapping peaks.}
\end{figure}

For the case of mass differences of comparable entity the energy peak
shapes could be largely overlapping, as sketched in the left panel
of Figure~\ref{fig:Two-examples-of-spectrum}. In this case the fitting
exercise is slightly more complicated because two peak shapes need
to be identified at once, however there is no fundamental difference
compared with the case described above. Of course the fit results
may degrade significantly because of the overlapping peaks shapes.
Ref. \cite{Agashe:2013ff} reports results for this type of ``merged''
peaks spectrum that we report in Figure~\ref{fig:Fits-overlapping}.
The identification of two peak positions is clearly still possible
despite they look like a single bump spectral shape. In fact it is
possible to use a $\chi^{2}$ fit to the spectrum with either one
or two peaks, modeled by eq.(\ref{eq:fitfunction}) or the sum of
two such functions, and choose what type of fit to perform based on
the $\chi^{2}$ value. In the cases studied by Ref. \cite{Agashe:2013ff}
the masses can be determined with accuracy slightly worse than those
attainable in the separated bumps case, but in any case the attainable
precision is  better than 10\%.

\begin{figure}
\begin{centering}
\includegraphics[width=0.88\linewidth]{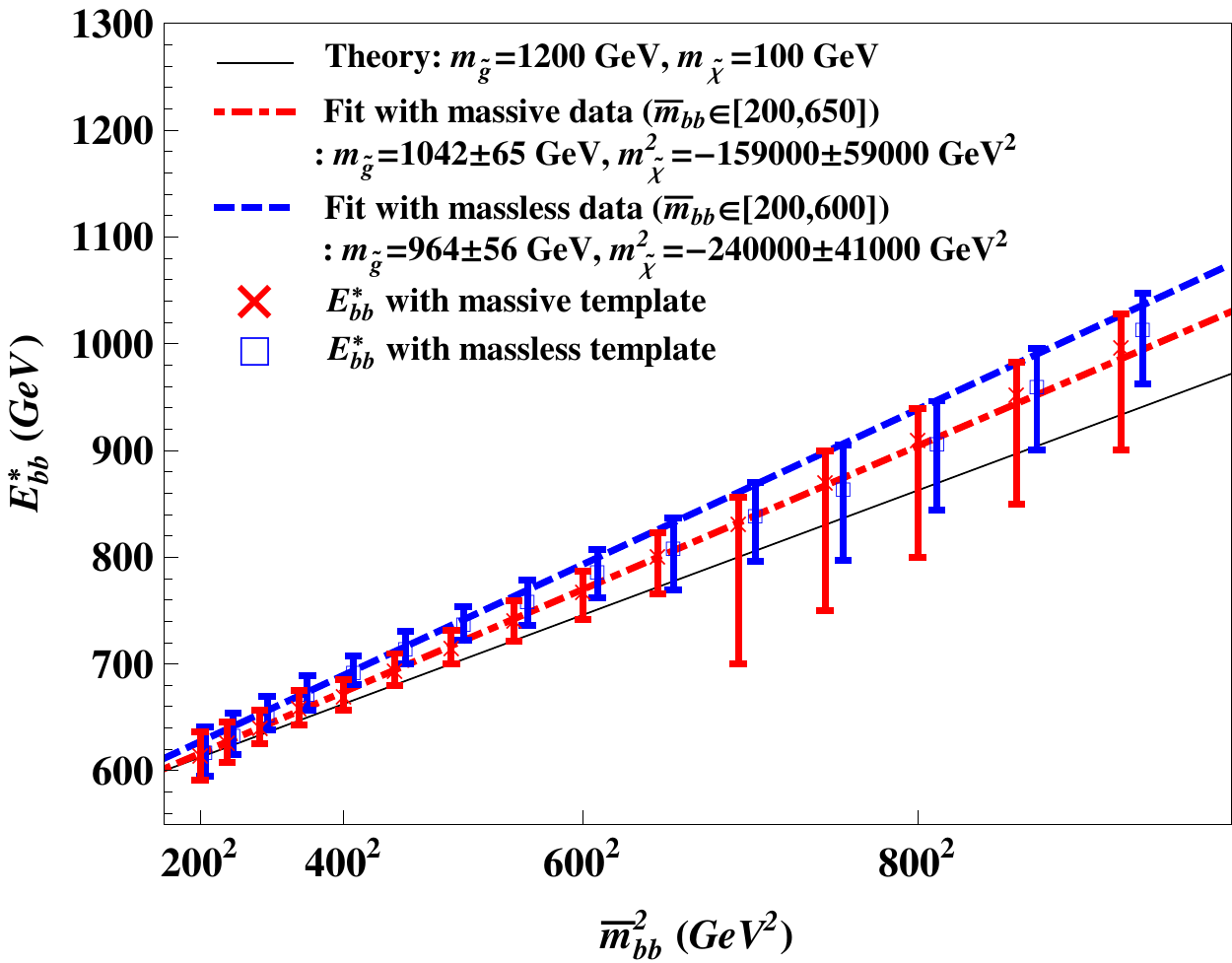}\caption{\label{fig:Energy-peak-3body}Energy peak extracted for different
slices of the three-body phase-space with approximatively constant
mass of the sub-system made of two $b$ quarks. Two lines represents
two different fit functions used to extract the peak value from the
data.}
\par\end{centering}
\end{figure}

All the above results have been derived in the context of two-body
decays. These can happen in cascades, as in the example discussed
above for gluino production and decay. However it is also possible
in new physics models that new particles do not have any allowed two-body
decay. In these cases the new particles might still decay via three-body
decays such as 
\[
\tilde{g}\to bb\chi^{0}\,.
\]
While this decay has the same final state as the above eq.(\ref{eq:cascade})
the underlying physics is different, because the no resonance exists
in the mass spectrum of any pair of final states. Despite this absence
of structure in the decay kinematics it is still possible to use energy
peaks analysis to extract information on the particles masses. In
fact any multi-body decay can be visualized as a (weighted) sum of
decays fewer bodies decays. For instance we could imagine the three-body
decay above to be described as a two body decay
\[
\tilde{g}\to B\chi^{0}
\]
 where the mass of the body $B$ changes in each recorded event within
the allowed limits $m_{B}\in[2m_{b},m_{\tilde{g}}-m_{\chi}]$. Decomposing
the decay process in this way we can deal independently with each
subset of the events with approximatively same value of $m_{B}$.
In practice we can observe the range of values than $m_{B}$ attains
in the data and divide it in sub-ranges $[m_{B}^{(i)}-\Delta,\,m_{B}^{(i)}+\Delta]$
in which we can consider $m_{B}$ to be constant and equal to $m_{B}^{(i)}$.
For each of the values $m_{B}^{(i)}$ an equation of the kind of eq.(\ref{eq:master-two-body})
will hold, because, within the error due to $\Delta$ being small
but not zero, the events effectively obey kinematical relations of
two-body decays. Therefore if we measure the energy peak of the subset
of events belonging to each of the $m_{B}^{(i)}$ values we can study
the relation 
\begin{equation}
E_{B^{(i)}}^{*}=\frac{m_{\tilde{g}}^{2}-m_{\chi}^{2}+\left(m_{B}^{(i)}\right)^{2}}{2m_{\tilde{g}}}\label{eq:3body2body}
\end{equation}
and fit the value of $m_{\tilde{g}}$ and $m_{\chi}$ from the several
values obtained for different $m_{B}^{(i)}$ choices. This strategy
to deal with three-body and multi-body decays has been investigated
in Ref.~\cite{Agashe:2015wj}. An example result for the determination
from data of the relation eq.(\ref{eq:3body2body}) is shown in Figure~\ref{fig:Energy-peak-3body}.
The figure also shows the comparison of two different choices for
the fitting function. The best performing one is a modification of
the parametrization introduced in eq.(\ref{eq:fitfunction}) that
Ref.~\cite{Agashe:2015wj} motivates to better take into account
the fact that $m_{B}^{(i)}$ can be large compared to $m_{\tilde{g}}-m_{\chi}$.
We refer the reader to Ref.~\cite{Agashe:2015wj} for further details
on the fit procedure adopted to extract $E_{B^{(i)}}^{*}$ for large
values of $m_{B}^{(i)}$. What we would like to highlight here, rather
than the details of the fit procedure, is the fact that phase-space
slicing allows to use ideas born for two-body decays in the context
of multi-body decays. 

As it is true in most cases, the information from energy peaks analysis
can be supplemented with the one coming form other measurements, such
as the end-point of the $m_{bb}$ spectrum. Combining all these information
Ref.~\cite{Agashe:2015wj} finds that the masses $m_{\tilde{g}}$
and $m_{\chi}$ can be determined with precision well below 10\%.
Given the non trivial interplay of fits and the role played by the
choice of the fitting function used to extract the peak values, in
the case of multi-body decays is possible to introduce noticeable
biases in the mass extraction. Ref.~\cite{Agashe:2015wj} finds that,
especially after a combination of the energy peaks results with the
results from invariant mass $m_{bb}$ analysis, these biases are much
mitigated.

The case of three-body decays, as we saw above, forces to deal with
the extraction of energy peaks from spectra of massive particles energies.
Once this becomes feasible a much larger range of decay processes
become tractable with energy peaks technique. An example of such decays
is the decay 
\[
\tilde{t}_{2}\to\tilde{t}_{1}Z^{0}
\]
which has been studied in Ref.~\cite{Agashe:2015nx}. In this work
greater details are given on how to model the shape of the energy
peak for massive particles. In particular it was shown that the peak
of the $Z$ boson energy distribution can be determined with precision
around or below 10\% and information of comparable accuracy on the
mass difference between the two particles $\tilde{t}_{2}$ and $\tilde{t}_{1}$
can be recovered looking solely at the $Z$ boson energy spectrum. 

In addition to the cases of pair production of new physics states,
energy peaks methods have been applied to single production of new
resonances whose mass can be measured in semi-invisible final states,
e.g. $pp\to G_{1}\to ZZ\to\nu\nu\mu^{+}\mu^{-}$ studied in Ref.~\cite{Chen:2014cr}
where the energy spectrum of the two muons, $E(\mu\mu)$ is used to
measure the mass of the intermediate resonance $G_{1}$. 

Before closing this section it is worth recalling a few applications
of related ideas on the use of energy distributions to measure particles
masses. In Ref.~\cite{Kawabata:2012kx} it has been considered how
a weighted average of the energy spectrum of a decay product can be
used to infer the mass of the decaying particle. The method outlined
differs from all the applications above for the fact that the peak
position is not central to the extraction of the mass of the decaying
particle, but rather the shape of the entire spectrum. Due to the
necessity of averaging on the whole spectrum, this kind of mass measurement
requires to either have fully inclusive data on the decay products
or, if part of the data is missing, to supplement the real data with
simulated one. This is usually an issue in high energy physics experiments
where, in a way or another, events are required to pass certain selections
in order to be useable for a mass measurement or simply to reject
backgrounds. The method has been studied at leading order in perturbation
theory for Higgs boson~\cite{Kawabata:2013fk} and top quark~\cite{Kawabata:2014ss}
mass measurement with encouraging results.

\section{Precision top quark mass measurement}

In the discussion above we have highlighted the several application
of energy spectrum peak measurements in the context of measurements
on putative new particles to be discovered at the Large Hadron Collider.
While these application are interesting \emph{per se}, it would be
even more interesting if these ideas could be applied on real data.
At the time of writing the Large Hadron Collider has not seen any
hint of new physics, therefore the energy peak method cannot be applied
yet on new physics mass measurements as outlined above. Nevertheless
there is an important piece of data on which it is interesting to
apply the energy peak method, that is the top quark data where energy
peaks techniques can help in the measurement of the mass of the top
quark.

It is important to stress that after more than 25 years of data collected
on top quarks at two different accelerators (TeVatron and the Large
Hadron Collider) we already know plenty about the top quark. Therefore
the applications discussed above of properties of energy peaks to
obtain mass measurements at 10\% precision are not necessarily interesting
when a single experiment can measure the top quark mass with precision
below 1\% using other techniques. However we have discussed in the
beginning section how one of the properties that make energy peaks
mass measurements interesting is that they make possible to carry
out a measurement while keeping the number of assumptions at a minimum.
For instance we have highlighted how energy peak relations such as
eq.~(\ref{eq:master-two-body}) hold irrespectively of the mechanism
or reaction used to produce the particle subject of mass measurement,
provided that the sample of particles that we observe is made of equal
populations of all possible particle polarizations. This fact is a
key to apply energy peaks techniques to a precision measurement such
as the measurement of the top quark mass.

In fact, even if we know the reaction for the production of the particle
of interest, in concrete cases we might know how to carry out detailed
calculations about this reaction only up to a certain order of perturbation
theory or neglecting certain effects which might go beyond the presently
developed modeling of fundamental interactions (\emph{e.g.} non-perturbative
or non-factorizable QCD effects). As a consequence of this unavoidable
presence of theory uncertainties, the energy peaks properties acquire
new interest in the context of mass measurements of Standard Model
particles. In fact, no matter what is the cause of mis-modeling of
production reactions, be it mis-modeling due to missing orders in
perturbation theory or absence of description of certain phenomena,
energy peaks predictions are expected to enjoy much milder theory
uncertainties than those of methods based on other observables. The
key of this resilience is the validity of results such as eq.~(\ref{eq:master-two-body})
for any two-body decay from a scalar particle or a sample of particles
with spin that populate evenly all polarization states.

In the specific case of top quark mass measurement all modern precise
mass measurements are obtained comparing some data with theory calculations~\cite{ATLAS:2014fj,Corcella:2017mkv,Cristinziani:2016qv,Cortiana:2015hp}.
The data is collected so far only in hadronic colliders and comes
from reactions such as $pp\to t\bar{t}$, which, due to hadronic nature
of the collision, should be more properly written $pp\to t\bar{t}+hadrons$.
In nowadays high energy physics practice the extra hadrons produced
together with the top quarks are clustered in sets of collimated particles
called \emph{jets}, therefore we will call this reaction 
\[
pp\to t\bar{t}+jets\,.
\]
The number of jets that is formed together with the top quarks roughly
corresponds to the number of orders in perturbation theory beyond
the lowest one at which one should have to carry out calculations
to describe this process with theoretical calculations. In addition,
processes that give rise to any number of jets contribute to the simplest
process $pp\to t\bar{t}$ when the jets do not carry large energies
or point in directions in which it is not possible to detect them.
These complications from the hadronic nature of the reaction that
produces the top quarks make difficult to obtain precision predictions
for any quantity measurable in top quark physics. In spite of all
this, calculations up to two orders beyond the leading one in QCD
have been carried out and it is nowadays possible to compute total
rates for top quark production with a theory uncertainty in the 5\%
ballpark~\cite{Czakon:2017uo,Czakon:2016ul,Czakon:2016ul,Czakon:2015owf,Czakon:2015nn}.
Furthermore it is possible to carry out calculations in which fixed
orders in perturbation theory are matched to calculations in which
large numbers of soft and collinear emissions of quanta are dealt
with in the more suitable parton shower picture, even with the inclusion
of non-resonant processes as well as resonant top quark pair production~\cite{Jezo:2016oj}.
After all these tremendous efforts to compute in ever more fine details
the production of top quarks at hadron colliders, theory errors on
the top quark mass determination are still not negligible and it has
become a tough problem to \emph{i}) quantify these uncertainties in
a reliable way \emph{ii}) combine measurements in a meaningful manner
when affected by such errors.

In view of the crucial role played by theoretical uncertainties in
the determination of the top quark mass from hadronic colliders the
use of energy peaks techniques may offer a new and complementary way
to look at these issues. In fact the insensitivity of the energy peak
relation eq.~(\ref{eq:master-two-body}) offers a chance for smaller
sensitivity to theory mis-modeling in extracting the top quark mass
from data. 

It is important to stress that in the context of a precision measurement
some of the assumptions that lead to the derivation of eq.~(\ref{eq:master-two-body})
may fail. For instance the necessity to impose selections on the data
to isolate top quark events from background events may ruin, even
if only slightly, the property of QCD to create samples of top quarks
that populate evenly all polarization states, which was a key fact
to extend the ``old'' results on decays of scalars to the case of
particles with spin. Furthermore the energy peak techniques are ideally
applied to two-body decays, so if one has radiative corrections to
the decay process,\emph{ e.g.} 
\begin{equation}
t\to bW+X\,,\label{eq:NLOdecay}
\end{equation}
then one has to review the validity of eq.~(\ref{eq:master-two-body})
to determine the top quark mass. These issues clearly need a dedicated
study, in which the dynamics of the Standard Model and in particular
of the QCD sector is used to model these effects and to assess their
effect on the determination of the top quark mass. To deal in a consistent
way with all these effects one can formulate a method to extract the
top quark mass by comparing data with different templates for energy
spectra corresponding to different values of the top quark mass. This
method, using spectra computed assuming the Standard Model dynamics
and including the first QCD corrections to production and decay of
the top quark \cite{MCFM63,Campbell:2012rt}, have been studied in
\cite{Agashe:2016xq}. It was found that theory errors due to missing
orders in the description of both the production and decay of top
quarks are rather small compared to those obtained for templates of
other observables (evaluated for instance in Ref.~\cite{Biswas:2010vq}).
Quite strikingly, this favorable comparison of theory uncertainties
is true even when one compares with observables that are rather close
to the energy, \emph{e.g.} the sum of the two energies of the $b$
quarks in each event studied in Ref.~\cite{Biswas:2010vq}. These
results confirm the possible utility of mass measurements, even precision
ones, based on single energy peaks techniques.

\begin{figure}
\begin{centering}
\includegraphics[width=0.49\linewidth]{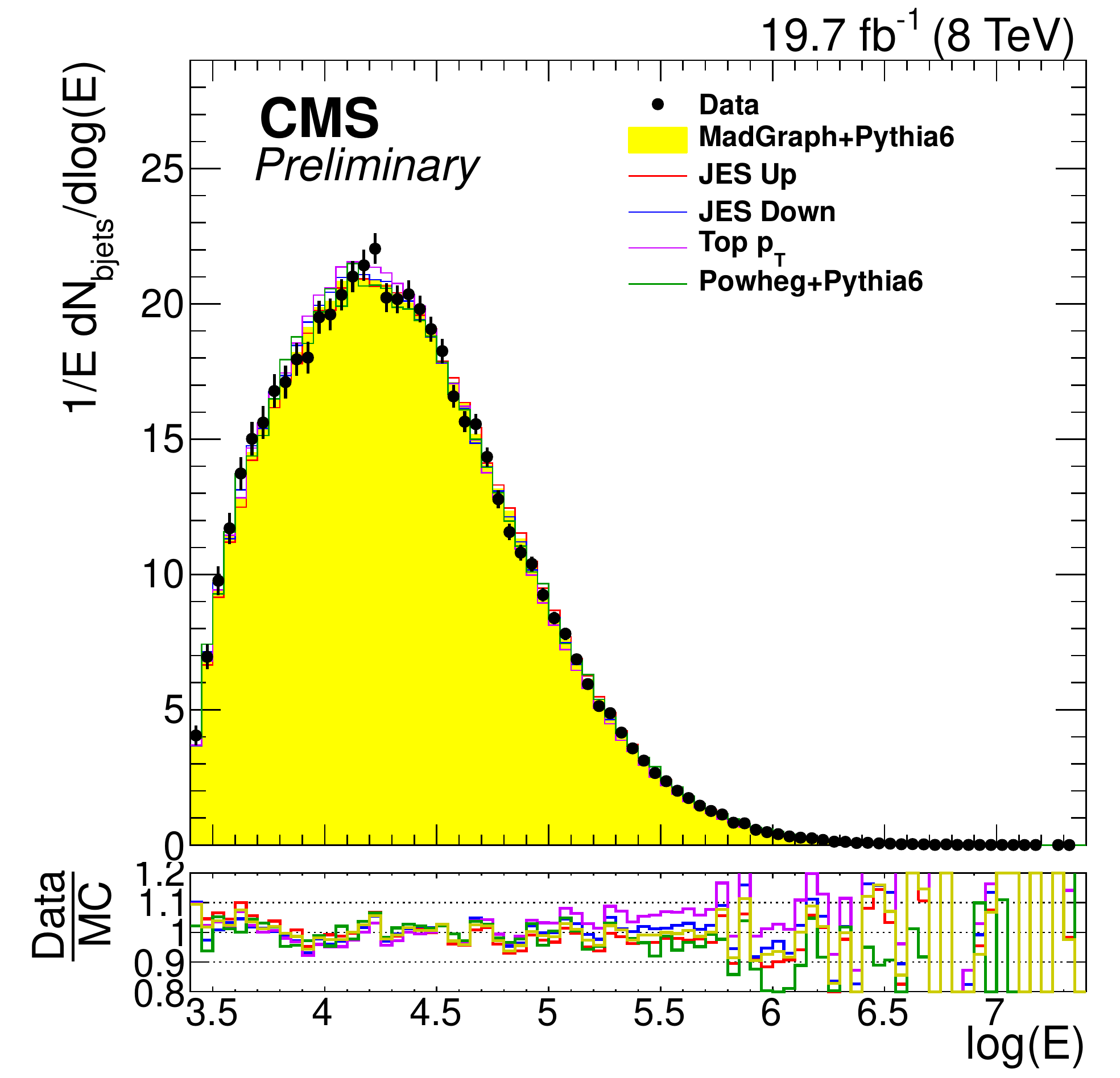}\includegraphics[width=0.49\linewidth]{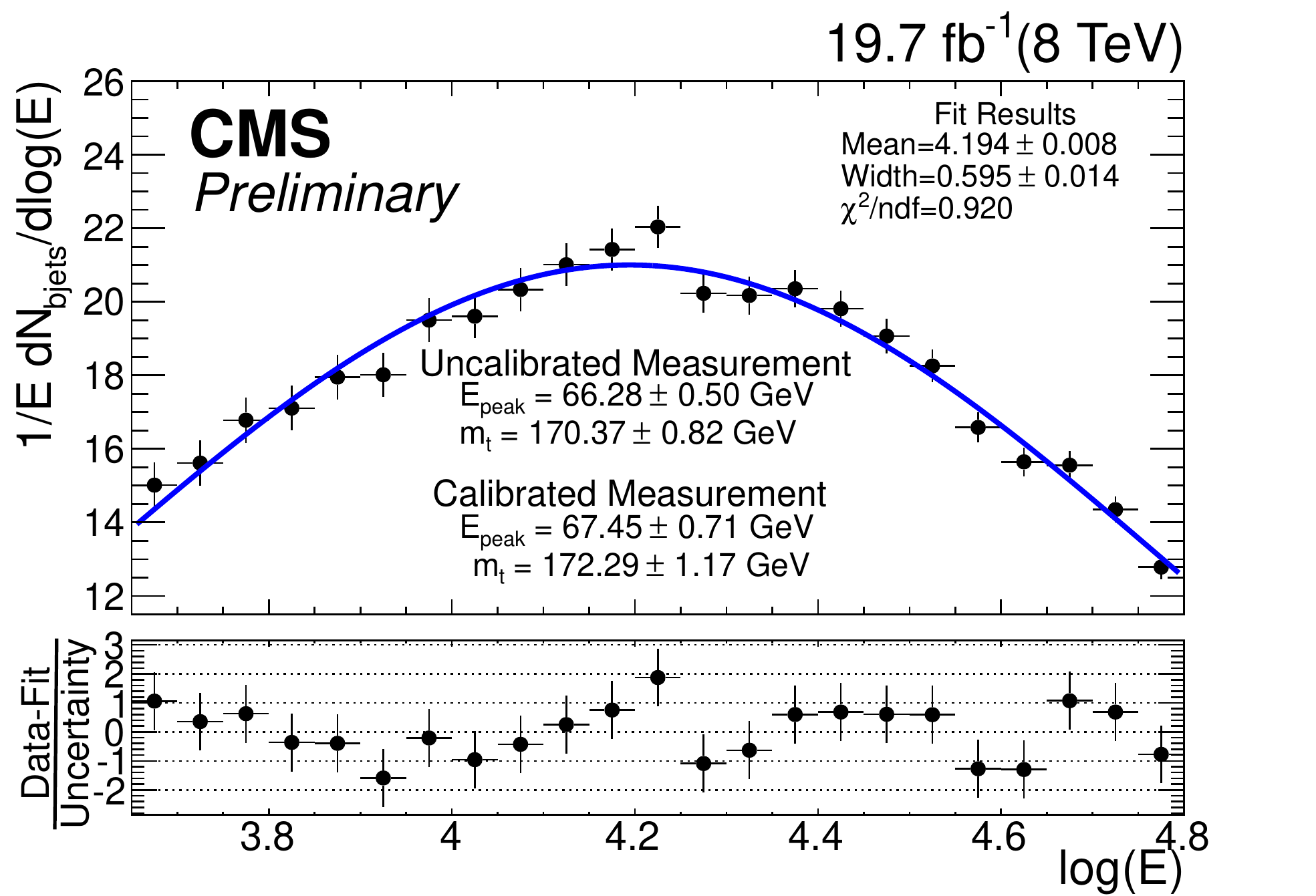}
\par\end{centering}
\caption{\label{fig:CMS-measurement}CMS measurement \cite{CMS-PAS-TOP-15-002}
of the $b$ jet energy measurement for the top quark mass measurement.}

\end{figure}

Furthermore it should be remarked that the CMS experiment has carried
out a preliminary measurement of the top quark mass using energy peaks
methods on early Large Hadron Collider data. The main result is reported
in Figure~\ref{fig:CMS-measurement}, that is the first ever public
measurement of an energy spectrum reported by CMS top quark physics
group to carry out a measurement. In Ref.~\cite{CMS-PAS-TOP-15-002}
the CMS experiment has used theory predictions at the lowest order
in perturbation theory to extract the top quark mass from the measured
$b$-jet energy spectrum and has obtained encouraging results. This
measurement has confirmed the estimate of the statistical error presented
in the original proposal of Ref.~\cite{Agashe:2013sw}. Furthermore
the theory error estimated by the CMS experiment is in line with expectations
for a result obtained from theory at the lowest order in perturbation
theory and leaves room for improvement once fixed-order and parton
shower improvements that have been developed in the meantime~\cite{Jezo:2016oj}
will be included in a future reiteration of this measurement.

\section{Dark Matter}

We have discussed how energy peak relations could be useful thanks
to the fact that they do not require to observe more than one decay
product for each decay. If such decay is a two-body decay we expect
eq.~(\ref{eq:master-two-body}) to hold up corrections that could
arise for instance from radiative corrections such as those we discussed
for the top quark decay eq.(\ref{eq:NLOdecay}). These corrections
are usually small, surely below 10\%, as they do not reach that level
even for a strongly interacting particle such as the top quark. Indeed
for a weakly interacting particles we expect eq.~(\ref{eq:master-two-body})
to hold very precisely.

One type of weakly interacting particles is very interesting in high
energy physics: the Dark Matter candidate particles. These are putative
particles that should explain a number of phenomena observed in the
Universe such as: the observed rotation speed of stars versus their
distance from the center of their galaxy, peaks in the power spectrum
of the Cosmic Microwave Background, gravitational lensing effects
of distant light sources and more (see Ref.~\cite{Bertone:2016lp,Bertone:2004pz}
for more details on the astrophysical and cosmological evidence for
Dark Matter). Such weakly interacting particles are often a piece
of a larger theory of physics beyond the Standard Model and are supposed
to fill present day Universe and have had a large impact on its evolution
ever since its formation. 

In order for Dark Matter particles to be present in our Universe from
its early times till today they are often charged under some special
symmetry that prevents them from decaying. Depending on the model
of new physics in which the Dark Matter particle is embedded this
symmetry might be a simple parity symmetry under which all particles
are either odd or even or some more complicated symmetry. Since ``even''
and ``odd'' are just two possible charges particles can have in
the simplest symmetry these are called $Z_{2}$ discrete symmetries.
Exactly like in arithmetics, two even charges make an even charge,
as well as two odd charges do, but one odd and one even charge make
an odd charge. Any interaction that is ``odd'', because it involves
an odd number of ``odd'' particles is forbidden. 

In most model of new physics a $Z_{2}$ symmetry is used to distinguish
the Standard Model particles, that are assigned to be ``even'',
and the new particles, that are assigned to be ``odd''. In such
a way any collision at particles colliders always has to produce a
pair of new physics particles and never one single particle. The case
seen above of gluino pair production eq.~(\ref{eq:guinopair}) is
an example of this very common feature in new physics phenomenology.
Similarly, when we considered the decay of the gluino in eq.~(\ref{eq:firststepgluino}),
it was a decay into one Standard Model particle, the $b$ quark, and
one new particle, the supersymmetric $\tilde{b}$ particle. For exactly
the same reason, the $Z_{2}$ symmetry, when we considered the decay
of the $\tilde{b}$ in eq.~(\ref{eq:secondstepgluino}) we had one
Standard model particle, a $b$ quark, and a new physics particle,
the $\chi$. In supersymmetric models that attempt to provide a Dark
Matter particle the $\chi$ particle is the lightest particle ``odd''
under the $Z_{2}$ symmetry. Therefore it is forbidden for $\chi$
to decay and, being electrically neutral, it can very well be a Dark
Matter candidate. 

The phenomenology of models for Dark Matter in which the Dark Matter
is stable because of a $Z_{2}$ symmetry can be all be captured by
the fact that the Dark Matter, when produced from a new particle decay,
is produced singly, that is to say a new particle $B$ decays 
\begin{equation}
B\to b\chi\,,\label{eq:Z2}
\end{equation}
where $b$ denotes a Standard Model particle and $\chi$ the Dark
Matter particle. In alternative models, in which a more complex symmetry
is used to prevent the Dark Matter from decaying, one might have more
than one Dark Matter per decay, for instance one could have 
\begin{equation}
B\to b\chi\chi\,,\label{eq:nonZ2}
\end{equation}
for suitable new physics particle $B$ and Standard Model particle
$b$. Since the $\chi$ particle is not directly observable, the visible
decay products of eq.~(\ref{eq:Z2}) and eq.~(\ref{eq:nonZ2}) at
first look are the same, which poses the problem of finding methods
to distinguish these two decay processes. The fact that energy peaks
formulae can be applied just observing one particle per decay suggests
that energy peaks methods can be helpful to answer this question.

Given the importance of the underlying symmetry that stabilizes the
Dark Matter, it is very important to explore methods to distinguish
the two classes of models that give rise to different decays eq.~(\ref{eq:Z2})
and eq.~(\ref{eq:nonZ2}). Ref.~\cite{Agashe:2012ij} explored how
to distinguish the two type of reactions leveraging the properties
of energy peaks. The idea is to observe the energy of $b$, the visible
decay product that appears in both kind of decays. If the peak of
the energy distribution matches with the prediction for a two body
decay eq.~(\ref{eq:master-two-body}) one could state with some confidence
level that the data suggests a $Z_{2}$ stabilization symmetry rather
than a more complicated symmetry. 

\begin{figure}

\includegraphics[width=0.49\linewidth]{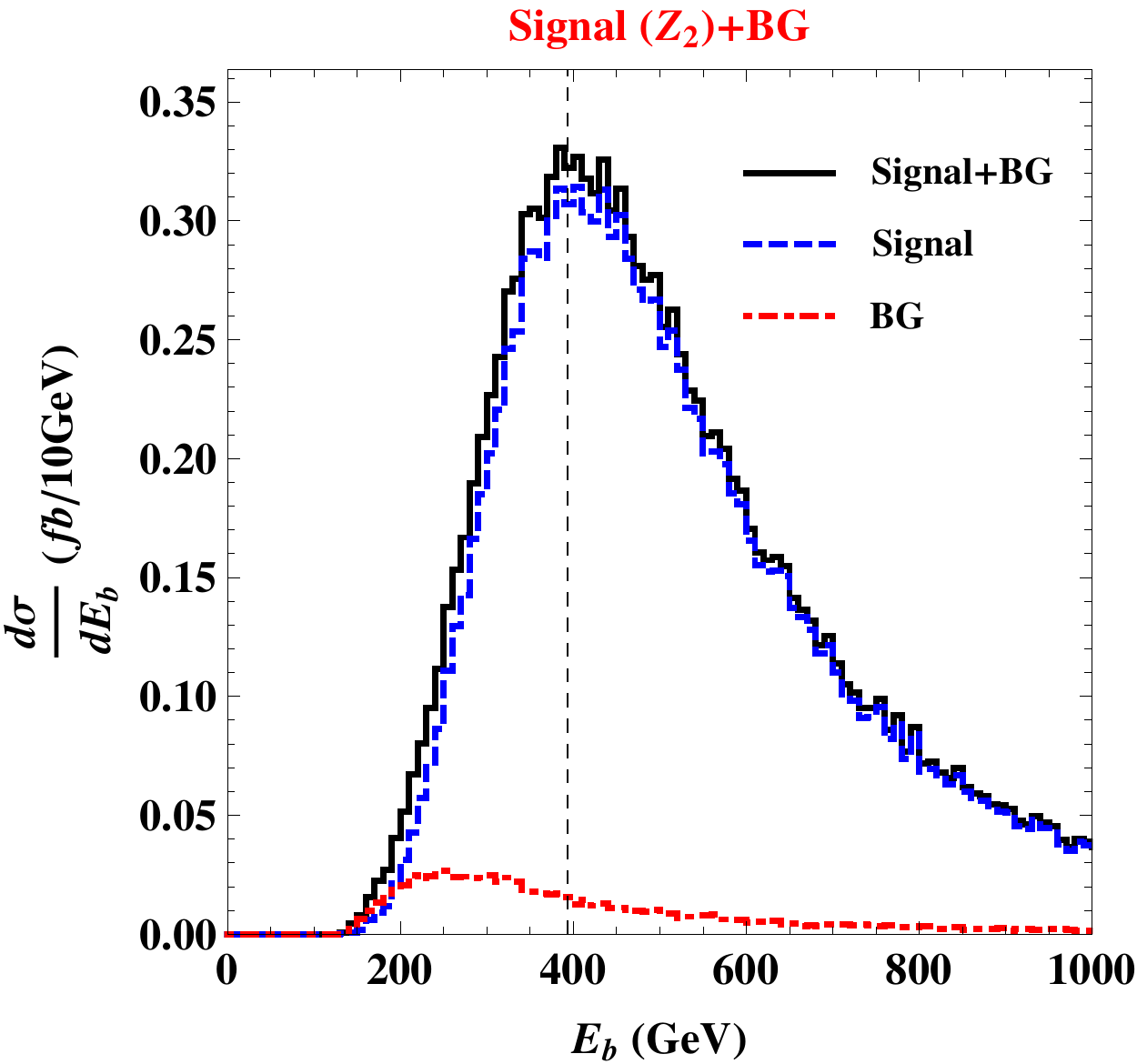}\includegraphics[width=0.49\linewidth]{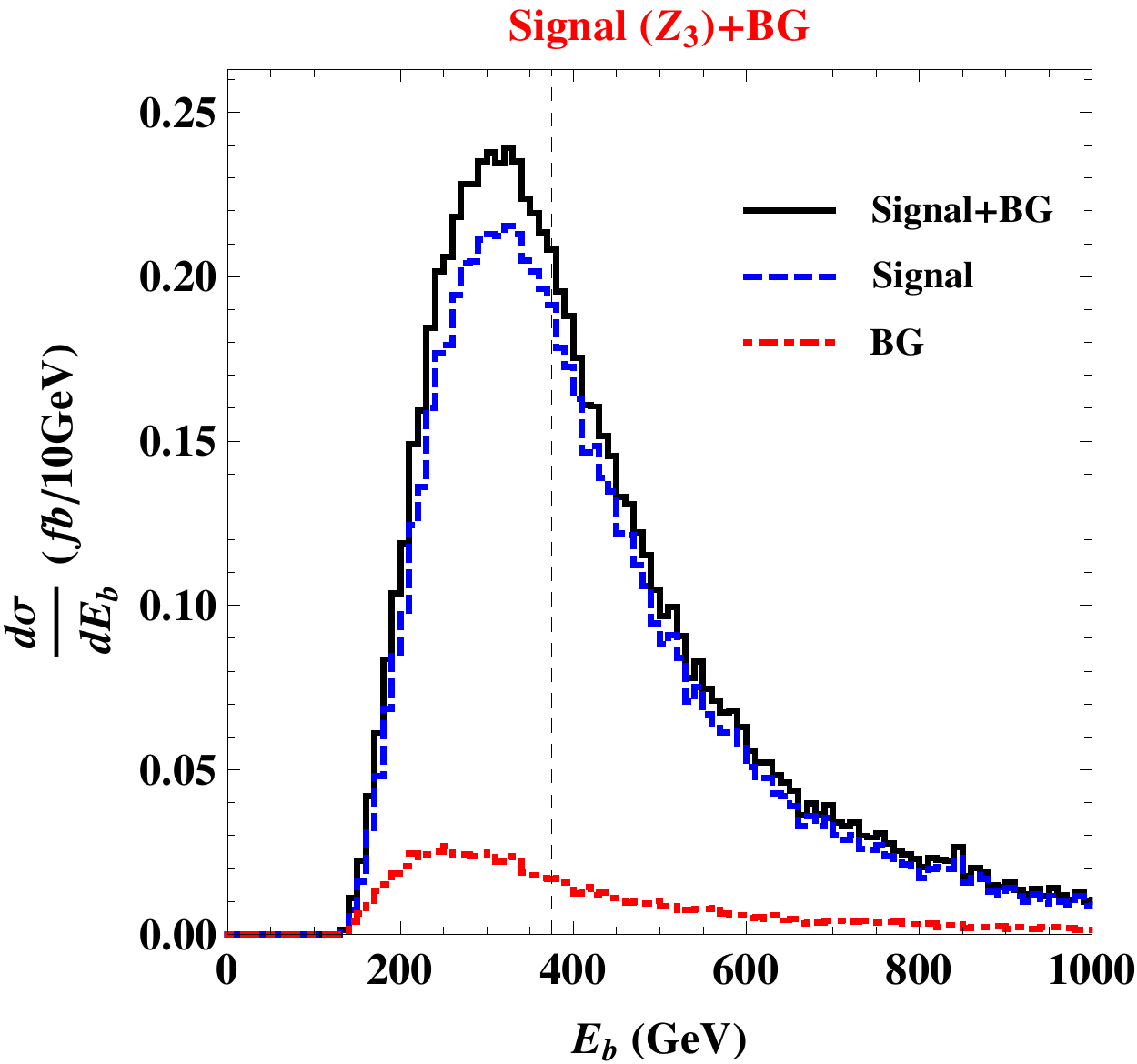}

\caption{\label{fig:Z2nonZ2} Example spectra for the $b$ particle energy
from a decay as in eq.~(\ref{eq:Z2}) and eq.~(\ref{eq:nonZ2})
in the left and right panel, respectively. The dashed lines are the
reference values extracted from the $m_{T2}$ analysis as described
in Ref.~\cite{Agashe:2012ij}.}

\end{figure}

It is important to stress that, in order to compute the correct value
for eq.~(\ref{eq:master-two-body}) one would need to know the masses
of the Dark Matter $\chi$ and of the heavier new physics particle
$B$, both of which might be poorly known. To surpass this difficulty
in Ref.~\cite{Agashe:2012ij} it has been pointed out that the value
of eq.~(\ref{eq:master-two-body}) relevant for a two-body decay
can be obtained from other distributions in particular from the end-point
of the variable $m_{T2}$ introduced in \cite{Barr:2003fj,Lester:1999et}.
The reference value from an $m_{T2}$ analysis can then be used to
check if the energy spectrum of the $b$ particle has a peak at that
value or at some lower values. An example of $b$ particle energy
spectra for the case of production of heavy fermions $B$ particles
and decay
\[
pp\to BB\to bb\chi\chi\text{ or }bb\chi\chi\chi\chi
\]
is shown in Figure~\ref{fig:Z2nonZ2}. In both panels the dashed
vertical line is the reference value obtained from an $m_{T2}$ analysis
which matches quite well on the left hand plot with the peak of the
distribution, while the right hand plot has a peak clearly shifted
with respect to the dashed line.

From the results of Ref.~\cite{Agashe:2012ij} it appears that if
new physics is discovered at colliders and the signals of new physics
contain invisible particles the energy peak method can be used to
count the number of invisible particles produced in each event, hence
probing very fundamental properties of the new physics theory such
as its symmetries and in particular the symmetry responsible for making
the Dark Matter stable.

We would like to remark that the properties of energy distributions
considered in these pages have also been recently applied in the context
of gamma rays astronomy for the characterization of Dark Matter signals.
Tell-tale tests of photon energy spectra from non-minimal Dark Matter
scenarios have been studied in Ref.~\cite{Boddy:2016aa}. In this
work it has been established that the features of the energy spectrum
of photons that arise from multi-particle Dark Matter sector, either
decaying or annihilating into photons or photons sources, can be revealed
with future gamma rays observatories. Furthermore in Ref.~\cite{Boddy:2017aa}
it was studied how features of the energy spectrum, such as the dependence
on $E/E^{*}+E^{*}/E$ that we pointed out to introduce eq.(\ref{eq:fitfunction}),
can be used to test the origin of present experiments excesses. 

Finally it is worth recalling that the development of eq.(\ref{eq:fitfunction})
to describe the energy spectrum of massless particles close to their
peak lead to the formulation of explanations of gamma rays excesses
in present experiment that significantly differ from the usually adopted
models of Dark Matter annihilation or decay. In the more standard
scenarios it is usually assumed that Dark Matter annihilates or decays
directly into SM states, \emph{e.g.} $\chi\chi\to\psi_{SM}\psi_{SM}$,
and the photons observed in the experiments are among the decay products
of the heavy SM states $\psi_{SM}$, that could be for instance SM
quarks, into stable SM states such as $e^{\pm},p,\bar{p,}\nu$, and
$\gamma$. Clearly, the connection between the observed gamma rays
spectrum and the properties of the Dark Matter sector is a complicated
one, because it may involve hadronization of quarks, details of the
formation of different species of QCD hadrons, bremsstrahlung of all
involved charged particles, and other aspects of the dynamics of the
Standard Model. At variance with the scenario of direct Dark Matter
annihilation or decay into SM states, in Refs.~\cite{Kim:2015lr,Kim:2015fe}
it has been proposed that bump-like features of the excess gamma rays
spectra can originate from multi-particle Dark Matter sectors through
decays $\chi^{\prime}\to\phi\chi$ followed by $\phi\to\gamma\gamma$
for mass $m_{\phi}$ around twice the energy of the peak region of
the excess. The spectral shape of this type of signal is in direct
connection with properties of the Dark Matter sector, hence allows
to test more directly the models of Dark Matter and in some cases
it has been shown to give better fits to the reported excess data~\cite{Kim:2015lr,Kim:2015fe}.

\section{Energy peaks: a ``one-prong'' Breit-Wigner?}

These results show that energy peaks are very useful tools to characterize
newly discovered particles, especially useful when we have experimentally
access to only one particle, or a subset, of the decay products of
a particle. We have seen how the energy distribution can be used to
probe the number of invisible particles in decays, \emph{e.g.} to
tell what symmetry makes Dark Matter a stable particle, or to measure
masses when some part of the kinematical information is carried away
by invisible particles and is not possible to draw a classic Breit-Wigner
distribution to identify a particle mass. 

Given the results obtained one might ask if an energy distribution
can be considered as some sort of ``one-prong'' Breit-Wigner distribution.
With this expression we mean to characterize energy peaks methods
as a generic substitute for invariant mass analysis in the cases in
which an invariant mass analysis is not possible for intrinsic properties
of the problem at hand, for instance the presence of invisible particles
as decay products which prevent from using four-momentum conservation.
At first this may look a too far fetched comparison. However it turns
out that is not at all inappropriate to put energy peaks next to well
known Breit-Wigner peaks. In fact both the existence of a peak in
a Breit-Wigner distribution and in an energy peak distribution rely
on the fact that an on-shell resonance can be thought as an intermediate
state of our reaction. For instance if we go back to top quark pair
production and we consider the off-shell contributions
\[
pp\to Wb\bar{t}
\]
neither the distribution of invariant mass $m_{Wb}$ nor that of the
energy of the $b$ quark will have a peak at $m_{t}$ or at the value
predicted by eq.~(\ref{eq:master-two-body}). Furthermore if one
considers extra radiation in the decay process, such as the correction
eq.~(\ref{eq:NLOdecay}) to the lowest order decay of the top quark,
the mass $m_{Wb}$ and the energy of the $b$ quark will again fail
to have peaks at $m_{t}$ and at the value predicted by eq.~(\ref{eq:master-two-body}),
respectively. 

The lesson learned here is that \emph{the same kind of hypotheses
lay behind the use of Breit-Wigner and energy peaks}. It should be
recalled however that for the energy peak relation to hold for a particle
with spin one needs to make sure that the observed sample of decays
has the same population of all the possible polarization states of
the particle. This may be trivially the case when we look at particles
without spin, but in general is something we have to check case by
case for each application of energy peaks methods. Production of particles
through strong and electromagnetic interaction, that conserve spatial
parity, can guarantee such even population of polarization states.
However one has to be careful to check that \emph{i}) experimental
effects such as event selection and \emph{ii}) subdominant reactions
mediated by weak interactions (that are violate spacial parity) do
not ruin the applicability of energy peaks results. For a fair comparison
it should be said that event selection can in principle ruin a Breit-Wigner
peak, but it is much harder to spoil such a peak than a peak in an
energy distribution.

Of course an energy peak analysis is always a second choice when a
Breit-Wigner analysis is possible, but we have seen that there are
plenty of interesting cases in which Breit-Wigner analysis simply
is unfeasible for absence of enough measurable particles (\emph{e.g.}
in cosmic rays studies, Dark Matter characterization and mass measurement
at colliders). Furthermore in cases in which a Breit-Wigner analysis
can be done there might still be interest in pursuing an energy peak
analysis to corroborate results and double-check uncertainty estimates
from the more powerful Breit-Wigner analysis, as we have seen for
the case of the precision top quark mass measurement.

\section{Outlook}

From the above discussion and the several results presented we have
learned that, despite the simplicity of the energy peak relation and
the hypotheses behind it, the use of energy peaks methods is very
powerful for certain problems in high energy particle physics. We
have seen that when particles decay into some combination of visible
and invisible particles it is possible to gain useful information
on the decay looking at peaks in energy distributions of the visible
particles.

In general if one has more than one visible particle per decay and
some number of invisible ones it is interesting to \emph{combine}
energy peaks techniques with more traditional ones based on invariant
masses, such as the end-point techniques, or methods based on transverse
momenta such as the $m_{T2}$ variable. In fact in most of the applications
we have discussed in this review the results obtained from energy
peaks have always been combined with other techniques. All in all
we have seen a nice degree of \emph{complementarity} between energy
peaks methods and other techniques more established in the high energy
particle physics practice.

We have reviewed the development of energy peaks techniques beyond
the historical results for spinless particles and we have extended
to the case of particles with spin. This leads to a new hypothesis
for the basic equation of energy peaks method to hold: the sample
of particles that is observed decay must populate evenly all possible
polarization states for eq.(\ref{eq:master-two-body}) to hold exactly.
Armed with this knowledge we have applied the energy peak method to
the measurement of masses of new physics particles that give rise
to complex final states containing invisible particles. We have studied
the determination of the mass of the gluino, the sbottom and the stop
supersymmetric particles in different contexts\cite{Agashe:2013ff}.
Along the way we have developed refinements of the original result
for decays into light or massless particle to cover the case of decays
into particles of non-negligible mass \cite{Agashe:2015nx}. We have
also generalized the idea of energy peaks to multi-body decays exploiting
phase-space slicing \cite{Agashe:2015wj} and the capability to deal
with massive decay products. All in all we have demonstrated the capability
of energy peaks methods to yield mass measurements with precision
at or below 10\% in a variety of contexts for new physics particles.

Furthermore we have applied the ideas of energy peak methods to distinguish
multi-body from two-body decays in the challenging case in which each
decay yields only one visible particle. This is precisely what would
be needed to tell which symmetry stabilizes the Dark Matter candidates
in new physics models and we have proven that energy peaks methods
can be used successfully to tell apart the simplest models of Dark
Matter particles from models with more complex symmetries.

We have also investigated the application of the energy peak method
to the precision measurement of particle masses. In the application
of energy peaks methods to precision mass measurements we have highlighted
the importance of radiative corrections to eq.(\ref{eq:master-two-body})
that should be taken into account when one seeks precision beyond
the lowest order in perturbation theory. We have discussed how, using
explicitly the dynamics of the Standard Model, these corrections can
be computed and indeed a method for the measurement of the top quark
mass that include these corrections has been devised in Ref.~\cite{Agashe:2016xq}.
The results in this work showed that precision measurements based
on energy peaks experience small theoretical uncertainties. From Ref.~\cite{Agashe:2016xq}
we can firmly conclude that that energy peak methods can be used for
a precision mass measurement once eq.(\ref{eq:master-two-body}) is
corrected according to the specific dynamics of the particles under
study.

Once we enter in a domain of precision physics, as it becomes inevitable
to use the dynamics of the Standard Model to carry out measurements,
it is possible to extend the use of energy peaks methods to other
measurements, such as that of the $W$ boson mass. In this case, being
the production of $W$ bosons inherently of electroweak nature, we
are not guaranteed at all that the $W$ bosons will be produced with
an even population of polarization states. However, using the dynamics
of the Standard Model, we can predict the population of each polarization
state. In general we expect small deviations from the predictions
of eq.(\ref{eq:master-two-body}) because of the roughly symmetric
collision environment at the Large Hadron Collider and of the small
velocity that most $W$ bosons have in the Drell-Yan reaction eq.(\ref{eq:DYW}).
The possibility to extract the $W$ boson mass from energy peaks has
not been studied in details yet, but it remains an interesting topic
for future work.

Within the contest of collider experiments we can envisage further
applications for energy peaks methods. For instance one could compare
the $b$ jet energy spectrum, and in particular the peak region, measured
in $pp$ collisions and in $p-ion$ and $ion-ion$ collisions recored
in the heavy ions physics runs of the Large Hadron Collider. From
these comparisons one can learn the interactions that heavy flavor
quarks, $b$ quarks in particular, have with the medium that is created
in the three types of collisions. This type of analysis would constitute
an interesting second look at effects due to the formation of quark-gluon
plasma such as the so-called ``jet quenching''~\cite{CasalderreySolana:2007pr,Apolinario:2016hi}.
Jet quenching is the dissipation of jets energy in collisions in which
two heavy ions $\mathcal{I}$ collide and give rise two jets, $\mathcal{I}\mathcal{I}\to jj$.
In principle these jets have equal and opposite momenta, and so they
do when measured ``in vacuum'', that is to say in relatively not
busy $pp$ collisions. However, in a heavy ion collision the nuclear
matter involved in the collision may give rise to a medium which tends
to alter the jet momentum by interactions with the quarks and gluons
that give rise to the jets. These interactions with the medium usually
result in an energy loss by the jet, hence the name ``quenching''. 

Thanks to experience gained on $b$ quarks energy spectra acquired
in top quark mass measurements it is in principle possible to study
this distribution in heavy ions collisions as well. Comparing the
$pp$ results to those for heavy ions, or studying how the distribution
changes when the heavy ions event activity changes, one can single
out the effects of the medium. Using the $b$ quarks from top decay
as hard probes of the medium one has the advantage of knowing quite
precisely the expected overall energy scale of the hard event and
in particular the expected energy peak position for $b$ quarks in
absence of medium effects. This may alleviate the uncertainties in
the study of dijet events in heavy ions collisions and constitute
a nice complement of the studies of jet quenching in events in which
a jet and a weak boson are produced, for instance $\mathcal{I}\mathcal{I}\to Zj$,
where the leptonic decay of the $Z$ boson offers a unit of measure
to check medium effects on the recoiling jet~\cite{CMS-Collaboration:2017aa}.

Leveraging the experience gained in the study of energy peaks in laboratory
experiments may also lead to novel uses of energy peaks techniques
outside high energy peaks laboratories. Looking in this direction
it is important to remark that in all the applications discussed above
for high energy physics it was critical to be able to extract the
peak position from a rather broad spectral shape (see for instance
the peak shape in the real data taken by the CMS experiment reported
in Figure~\ref{fig:CMS-measurement}). The capability to extract
this peak arises from a very good modeling of the peak shape and to
a certain extent of the tails of spectrum. The fact that this modeling
has been validated on data from a laboratory experiment with rather
small uncertainties might enable new applications of energy peaks.
For instance it appears feasible to look for traces of mesons heavier
than the $\pi^{0}$ in cosmic rays, \emph{e.g.} in satellite-born
experiments such as FERMI or the future ComPair~\cite{Moiseev:2015lva}
and e-ASTROGAM~\cite{Tatischeff:2016ykb} experiments that could
be sensitive to $\eta$ meson production in supernova remnants. The
signal from these $\eta$ mesons would be hidden underneath that of
the more abundant $\pi^{0}$ claimed by FERMI in Ref.~\cite{Ackermann:2013oq},
therefore it is very important to be able to model the whole shape
of the $\pi^{0}$ peak up to photon energy values where the $\eta$
peak might be visible. Observing such a $\eta$ meson signal would
put on firmer ground the hadronic origin of a major part of the cosmic
radiation and would be a nice come-back of energy peaks methods to
cosmic rays physics.

\section*{Acknowledgements}

All the knowledge and results reported here on energy peaks are the
result of many beautiful discussions and collaborations with Kaustubh
Agashe and Doojin Kim, whom I would like to thank very much. Furthermore
I would like to thank collaborators in research projects Sungwoo Hong
and Kyle Wardlow. Finally I would like to thank the Top LHC Working
Group and in particular Michelangelo Mangano, Martijn Mulders, Pedro
Silva, Benjamin Stieger for many discussions about the application
of these ideas to the concrete case of the measurement of the top
quark mass. The typography of this work is inspired by Ref.\cite{Gupta:2013mb}.

\bibliographystyle{utphys}
\bibliography{/Users/roberto/Dropbox/BibReader/mybibliography}

\end{document}